\documentclass{PoS}

\newcommand{\lsim}{ {\
\lower-1.2pt\vbox{\hbox{\rlap{$<$}\lower5pt\vbox{\hbox{$\sim$}}}}\ } }
\newcommand{\gsim}{ {\
\lower-1.2pt\vbox{\hbox{\rlap{$>$}\lower5pt\vbox{\hbox{$\sim$}}}}\ } }

\def\er#1#2{\relax\ifmmode{}^{+#1}_{-#2}\else$^{+#1}_{-#2}$\fi}
\def\erparen#1#2{\relax\ifmmode{}(^{#1}_{#2})\else$(^{#1}_{#2})$\fi}
\def~{\ifmmode\phantom{0}\else\penalty10000\ \fi}

\def\ev{\mathrm{e\kern-0.1em V}}
\def\kev{\mathrm{ke\kern-0.1em V}}
\def\mev{\mathrm{Me\kern-0.1em V}}
\def\gev{\mathrm{Ge\kern-0.1em V}}
\def\tev{\mathrm{Te\kern-0.1em V}}

\def\n#1e#2n{{#1}\times 10^{#2}}

\def\bea{\begin{eqnarray}}
\def\eea{\end{eqnarray}}

\def\ods2{\mathcal{O}_{\Delta S=2}}
\def\zds2{Z_{\Delta S=2}}

\makeatletter
\def\slash#1{{\mathpalette\c@ncel{#1}}} 
\def\big#1{{\hbox{$\left#1\vbox to1.012\ht\strutbox{}\right.\n@space$}}}
\def\Big#1{{\hbox{$\left#1\vbox to1.369\ht\strutbox{}\right.\n@space$}}}
\def\bigg#1{{\hbox{$\left#1\vbox to1.726\ht\strutbox{}\right.\n@space$}}}
\def\Bigg#1{{\hbox{$\left#1\vbox
to2.083\ht\strutbox{}\right.\n@space$}}}
\makeatother

\title{Computing the nucleon sigma terms at the physical point}

\ShortTitle{Computing the nucleon sigma terms at the physical point}

\author{\speaker{Christian Torrero} for the Budapest-Marseille-Wuppertal collaboration\\
        CNRS, Aix-Marseille Universit\'e, Universit\'e de Toulon,\\ Centre de Physique Th\'eorique (CPT, UMR 7332),\\ F-13288 Marseille, France\\
        E-mail: \email{Christian.Torrero@cpt.univ-mrs.fr}}

\abstract{Nucleon sigma terms are quantities that play an important role in various areas: among others, they connect the pion-nucleon and the kaon-nucleon amplitudes to the hadron spectrum and they are also relevant for the direct detection of Dark Matter. We present preliminary results for the up-down and strange sigma terms obtained from $N_f=2+1$ lattice simulations that are performed at five lattice spacings and for pion masses all the way down to its physical value.}

\FullConference{The 32nd International Symposium on Lattice Field Theory,\\
		23-28 June, 2014\\
		Columbia University New York, NY}

\begin{document}

\section{Introduction}

\begin{figure}
\hspace*{0.0cm}\includegraphics[width=7.2cm, height=7.4cm]{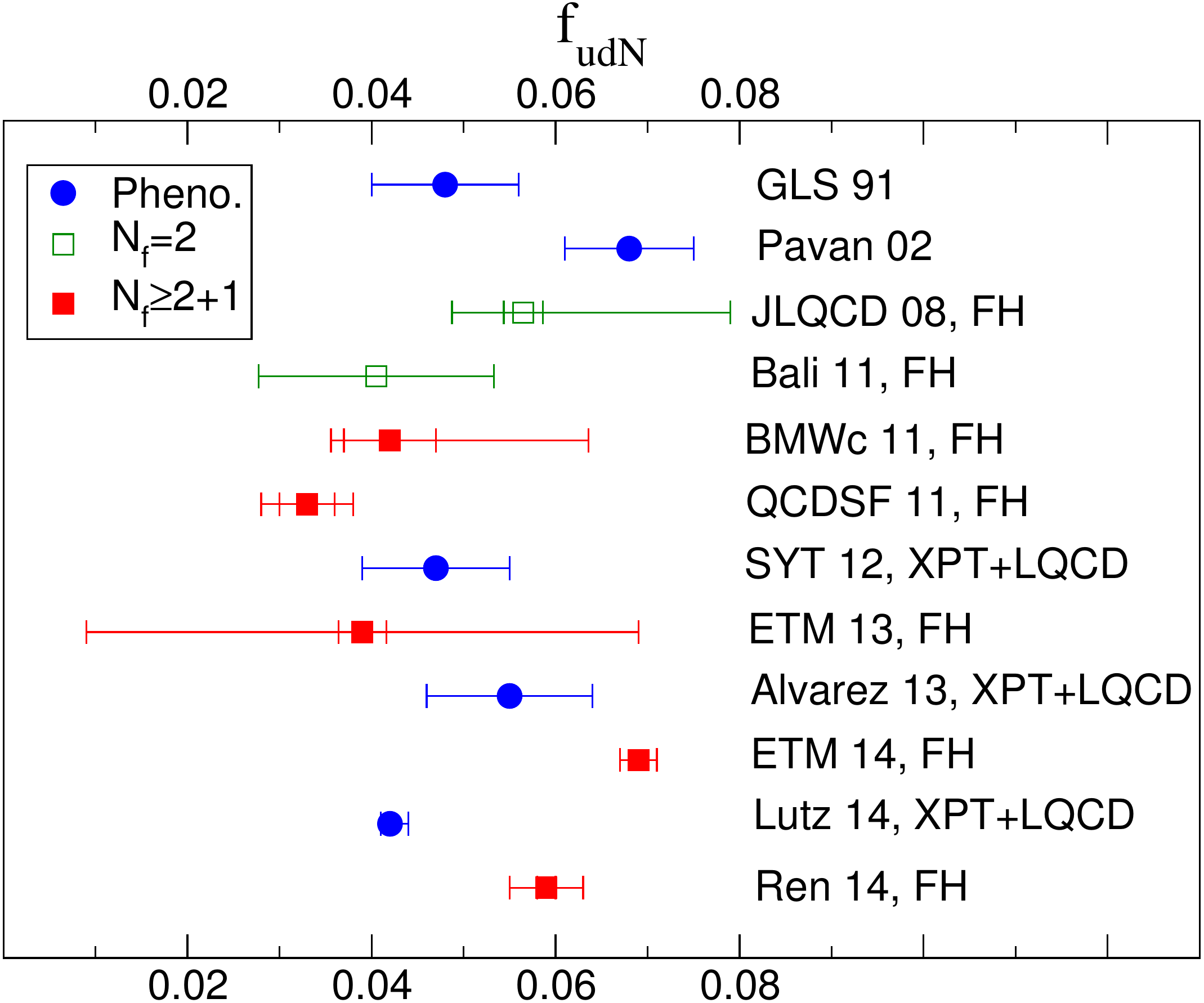}\hspace*{0.5cm}
\includegraphics[width=7.2cm, height=7.4cm]{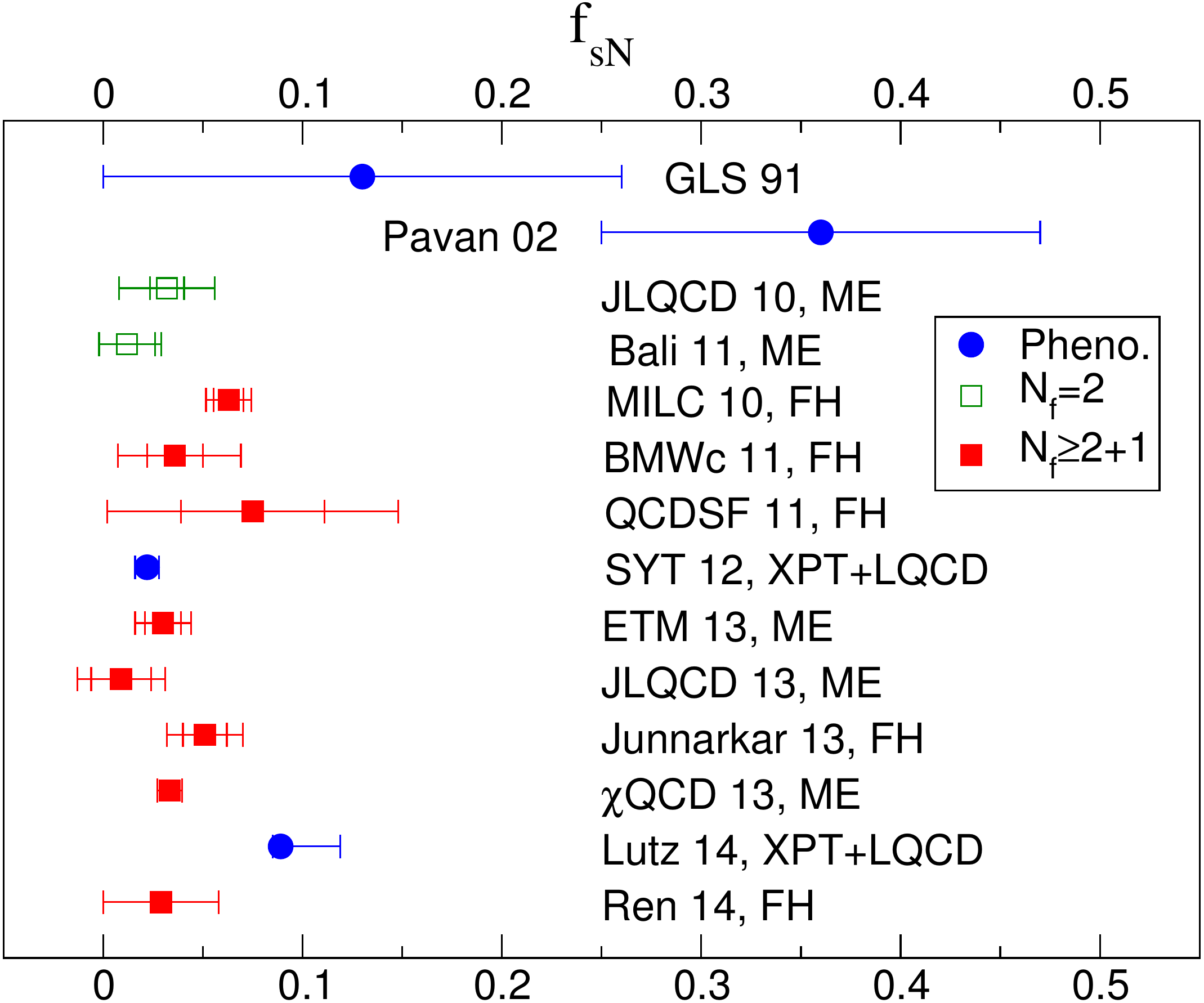}
\vspace*{-0.3cm}
\caption{Comparisons among phenomenology and lattice QCD computations of the nucleon up-down and strange quark content. See e.g. Refs. \cite{Alexandrou14,Ren14} and references
therein for details.}
\label{Fig. 1}
\vspace*{0.5cm}
\end{figure}

The nucleon sigma terms and the associated quark contents are observables of great interest given their relation to the quark mass ratio $m_{ud}/m_s$ and $\pi-N$ and $K-N$ scattering. They also have considerable importance in the direct detection of dark matter since they play a crucial role in the Dark Matter-quark coupling. Nucleon sigma terms are defined as\footnote{Note that there are different conventions for the definition of $\sigma_{\bar{\!s}s N}$: the one we adopted here is such that the two sigma terms are equal at the $SU(3)$ symmetric point.}   

\vspace*{-0.3cm}
\begin{eqnarray}
\sigma_{\!\pi N} &\equiv& m_{ud}\langle N\vert \bar uu+\bar dd\vert N\rangle \ ,\nonumber\\
\sigma_{\bar{\!s}s N} &\equiv& 2m_s\langle N\vert \bar ss\vert N\rangle\ ,
\end{eqnarray}
\vspace*{-0.2cm}

\noindent while the nucleon quark contents read

\vspace*{-0.3cm}
\begin{eqnarray}
f_{udN} &=& \frac{m_{ud}\langle N\vert \bar uu+\bar dd\vert N\rangle}{M_N} = \frac{\sigma_{\!\pi N}}{M_N}\ ,\nonumber\\
f_{sN} &=& \frac{m_s\langle N\vert \bar ss\vert N\rangle}{M_N} = \frac{\sigma_{\bar{\!s}s N}}{2M_N}\ .
\end{eqnarray}
\vspace*{-0.2cm}

Even though they are not directly accessible to experiment, they can be computed through phenomenology with results which, however, do not agree and are plagued with large uncertainties~\cite{Gasser91,Pavan02}. An alternative method to determine them consists in computing them using lattice QCD simulations: this strategy has been followed in recent years by different groups (for a collection of results, see Fig. $\!\!\!1$ and 
\cite{Alexandrou14,Ren14} with references therein), though computations have often featured model assumptions whose impact on final estimates cannot be fully assessed.\\ 
\indent In the framework of the Budapest-Marseille-Wuppertal collaboration, the present work aims at a first-principle computation of sigma terms with a complete error analysis: the initial approach is based on the well-known Feynman-Hellman theorem relating nucleon sigma terms to the quark dependence of the nucleon mass, i.e.,

\vspace*{-0.4cm}
\begin{eqnarray}
\sigma_{\!\pi N} &=& \left. m_{ud}\frac{\partial M_N}{\partial m_{ud}}\right|_{\Phi}\ ,\nonumber\\ 
&&\nonumber\\
\sigma_{\bar{\!s}s N} &=& 2\!\left.m_{s}\frac{\partial M_N}{\partial m_{s}}\right|_{\Phi} \ , 
\end{eqnarray}
\vspace*{-0.2cm}

\noindent where derivatives have to be computed at the \emph{physical point} ($\Phi$). With respect to a direct computation, the main disadvantage of this method is due to the fact that derivatives above are small (in particular for the $s$ case); however, such an approach demands computing 2-point functions only, avoids any challenging disconnected contribution and its underlying renormalization pattern is much less involved: these advantages make the strategy worth following.

\section{Simulation and analysis details}

An exhaustive description of the algorithm and the simulation details can be found in~\cite{Durr10}. Here it suffices to know that simulations feature tree-level improved Symanzik gauge action and $N_f = 2+1$ clover-improved Wilson action with $2$ levels of HEX link smearing. The analysis has been carried out on $47$ ensembles corresponding to about $13000$ overall configurations with five values of the lattice spacing $a$ in between $0.054$ fm and $0.116$ fm, pion mass down to $120$ MeV and box sizes up to $6$ fm.\\
\indent In order to apply the Feynman-Hellman theorem, we compute the lattice nucleon mass $\widehat{M_N}$\footnote{In what follows, quantities in lattice units will be labelled with a hat.} together with the lattice mass $\ \!\widehat{M_{\Omega}}$ of particle $\ \!\Omega$, the latter being used to set the scale. However, compared to a previous study~\cite{Durr12} where quark masses dependences of hadron masses were studied in terms of $M_{\pi}^2$ and $M_{K\chi}^2=M_K^2-M_{\pi}^2/2$, we now perform an expansion with respect to quark masses $m_{ud}$ and $m_s$ directly, thus no longer relying on leading order $SU(3)$ $\chi$PT relations in Eq. (1.3). However, since the physical value of quark masses cannot be measured experimentally, two more particle masses have to be fitted together with $\widehat{M_N}$ and $\widehat{M_{\Omega}}$ in order to fix $m_{ud}^{(\Phi)}$ and $m_{s}^{(\Phi)}$. The masses that we opted for are $\widehat{M_{\pi}}$ and $\widehat{M_{K\chi}}$.\\
\indent Altogether, the fit is made up of $4$ functional forms reading: 

\begin{eqnarray}
\widehat{M_{\Omega}} &=& aM_{\Omega}^{(\Phi)}\Bigg\{1+ c_{\Omega,ud,1}\Bigg[\frac{\widehat{m_{ud}}Z_s^{-1}(\beta)}{a(1+c_{ud}a^2)M_{\Omega}^{(\Phi)}}-\frac{m_{ud}^{(\Phi)}}{M_{\Omega}^{(\Phi)}}\Bigg] +c_{\Omega,s,1}\Bigg[\frac{\widehat{m_{s}}Z_s^{-1}(\beta)}{a(1+c_{s}a^2)M_{\Omega}^{(\Phi)}}-\frac{m_{s}^{(\Phi)}}{M_{\Omega}^{(\Phi)}}\Bigg]\Bigg\} \ ,\nonumber\\
&&\nonumber\\
&&\nonumber\\
\widehat{M_{\pi}} &=& a\bigg\{M_{\pi}^{(\Phi)} + \sum_{i=1}^3c_{\pi\!,ud,i}\bigg[\frac{\widehat{m_{ud}}Z_s^{-1}(\beta)}{a(1+c_{ud}a^2)}-m_{ud}^{(\Phi)}\bigg]^{\!i}+c_{\pi\!,s,1}\bigg[\frac{\widehat{m_{s}}Z_s^{-1}(\beta)}{a(1+c_{s}a^2)}-m_{s}^{(\Phi)}\bigg]\bigg\} \ ,\nonumber\\
&&\nonumber\\
&&\nonumber\\
\widehat{M_{K\chi}} &=& a\bigg\{M_{K\chi}^{(\Phi)} + c_{K\!,ud,1}\bigg[\frac{\widehat{m_{ud}}Z_s^{-1}(\beta)}{a(1+c_{ud}a^2)}-m_{ud}^{(\Phi)}\bigg]
+\sum_{i=1}^2c_{K\!,s,i}\bigg[\frac{\widehat{m_{s}}Z_s^{-1}(\beta)}{a(1+c_{s}a^2)}-m_{s}^{(\Phi)}\bigg]^{\!i}\ \!\bigg\} \ ,\nonumber\\
&&\nonumber\\
&&\nonumber\\
\widehat{M_{N}} &=& a\bigg\{M_{N}^{(\Phi)} + \sum_{i=1}^2c_{N\!,ud,i}\bigg[\frac{\widehat{m_{ud}}Z_s^{-1}(\beta)}{a(1+c_{ud}a^2)}-m_{ud}^{(\Phi)}\bigg]^{\!i}
+c_{N\!,s,1}\bigg[\frac{\widehat{m_{s}}Z_s^{-1}(\beta)}{a(1+c_{s}a^2)}-m_{s}^{(\Phi)}\bigg]\ \!\bigg\} \ ,
\end{eqnarray}
\vspace*{0.1cm}

\noindent where experimental input is represented by $M_{\Omega}^{(\Phi)}$, $M_{\pi}^{(\Phi)}$ and $M_{K\chi}^{(\Phi)}$ while quantities in lattice units are the numerical input. Note that there is one parameter $a$ for each value of $\beta$. With the notation above and thanks to the Feynman-Hellman theorem, nucleon sigma terms are related to fit parameters through

\begin{equation}
\sigma_{\!\pi N} = m_{ud}^{(\Phi)}c_{N\!,ud,1}\ , \ \ \ \ \ \ \  \sigma_{\bar{\!s}s N} = 2m_{s}^{(\Phi)}c_{N\!,s,1}\ ,
\end{equation}
\vspace*{-0.18cm}

\noindent while quark contents are given by

\begin{equation}
f_{udN} = m_{ud}^{(\Phi)}\frac{c_{N\!,ud,1}}{M_{N}^{(\Phi)}}\ , \ \ \ \ \ \ \ f_{sN} = m_{s}^{(\Phi)}\frac{c_{N\!,s,1}}{M_{N}^{(\Phi)}}\ .
\end{equation} 
\vspace*{-0.05cm}

\indent Input masses $\widehat{M_{\Omega}}$, $\widehat{M_{\pi}}$, $\widehat{M_{K\chi}}$ and $\widehat{M_N}$ have been computed with a standard procedure, i.e., by fitting the asymptotic behaviour of time correlators, 
while lattice quark masses $\widehat{m_{ud}}$ and $\widehat{m_s}$ have been obtained through a \emph{ratio-difference method} and have been renormalized by means of the renormalization constant $Z_s$ computed non-perturbatively as done in \cite{Durr10}.






\indent The four hadron masses appearing in Eqs. (2.1) are fitted simultaneously, i.e., they share the same fit parameters with the obvious exception of coefficients $c_{X\!,ud,j}$ and $c_{X,s,j}$. Fit parameters $c=\{a,m_{ud}^{(\phi)},m_{s}^{(\phi)},\ldots\}$ of functions $f^{(i)}(c,x)$ --- with $i=1,2,3,4$ and $x=\{\widehat{m_{ud}},\widehat{m_s}\}$ --- are determined by minimizing a $\chi^2$ function defined as

\vspace*{-0.01cm}
\begin{equation}
\chi^2 = V^{T}C^{-1}V\ ,
\end{equation} 
\vspace*{-0.3cm}

\noindent $C$ being the covariance matrix associated to the entries of the column vector $V$ whose structure reads

\vspace*{-0.24cm}
\begin{equation}
V = (y^{(1)}_1 -\ f^{(1)}(c,x_1)\ ,\ \ldots\ ,\ y^{(4)}_n-\ f^{(4)}(c,x_n)\ ,\ x_1 - q_1\ ,\ x_2 - q_2\ ,\ \ldots\ ,\ x_n - q_n)\ ,
\end{equation}
\vspace*{-0.3cm}

\noindent where $q_i$ is the value obtained for variable $x_i$ in simulation $i$. The entries of matrix $C$ have been estimated by means of a bootstrap procedure with $2000$ samples. All fits are correlated.\\
\indent In order to estimate the systematic uncertainties on results, different strategies have been considered for the analysis:

\begin{itemize}

\item choosing two different time intervals for the asymptotic behaviour of time correlators;

\item pruning the data with two cuts in the pion mass (at $380\ \mev$\ and $480\ \mev$);

\item taking into account six different procedures for computing $Z_S$ as in~\cite{Durr10};

\item relying on various $\chi$PT-inspired fitting functions for mesons;

\item allowing for different cutoff effects, i.e.

\vspace*{-0.1cm}
\begin{equation}
\frac{am_{ud}Z_s^{-1}(\beta)}{a(1+d_{ud}a^2)} \longrightarrow \frac{am_{ud}Z_s^{-1}(\beta)}{a(1+d_{ud}a\alpha_s(a))}\ ,
\end{equation}
\vspace*{-0.25cm}

\noindent $\alpha_s(a)$ being the strong-coupling constant at scale $a$. 

\end{itemize}

\noindent This results in $2\cdot2\cdot6\cdot2\cdot2 = 96$ fitting strategies altogether. In principle, also finite volume corrections should be taken into account but they are negligible compared to other uncertainties here, since $M_{\pi}L\gtrsim4$ in our volumes.\\
\indent The systematic error on each quantity is then evaluated by means of the \emph{Akaike Information Criterion} \cite{Borsanyi14}, i.e. for the $i^{\ \!th}$ analysis procedure the AIC value AIC$_i$ 

\begin{equation}
AIC_i = 2k_i - 2\ln(L_i)\ ,
\end{equation} 
\vspace*{-0.3cm}

\noindent is computed, $k_i$ being the number of fit parameters and $L_i$ the maximized value of the likelihood function. The statistical weight $\omega_i$ of the procedure is then given by

\begin{equation}
\omega_i = e^{-(AIC_i-AIC_{min})/2}\ ,
\end{equation} 
\vspace*{-0.2cm}

\noindent $AIC_{min}$ being the smallest of the $AIC_i$'s. The mean value and systematic error of a generic fit parameter $c_j$ are obtained by computing, respectively, the AIC-weighted mean and standard deviation of the values of $c_j$ resulting from the different analysis procedures. The bootstrap error on the mean provides the statistical uncertainty.

\begin{figure}
\includegraphics[width=0.85\textwidth]{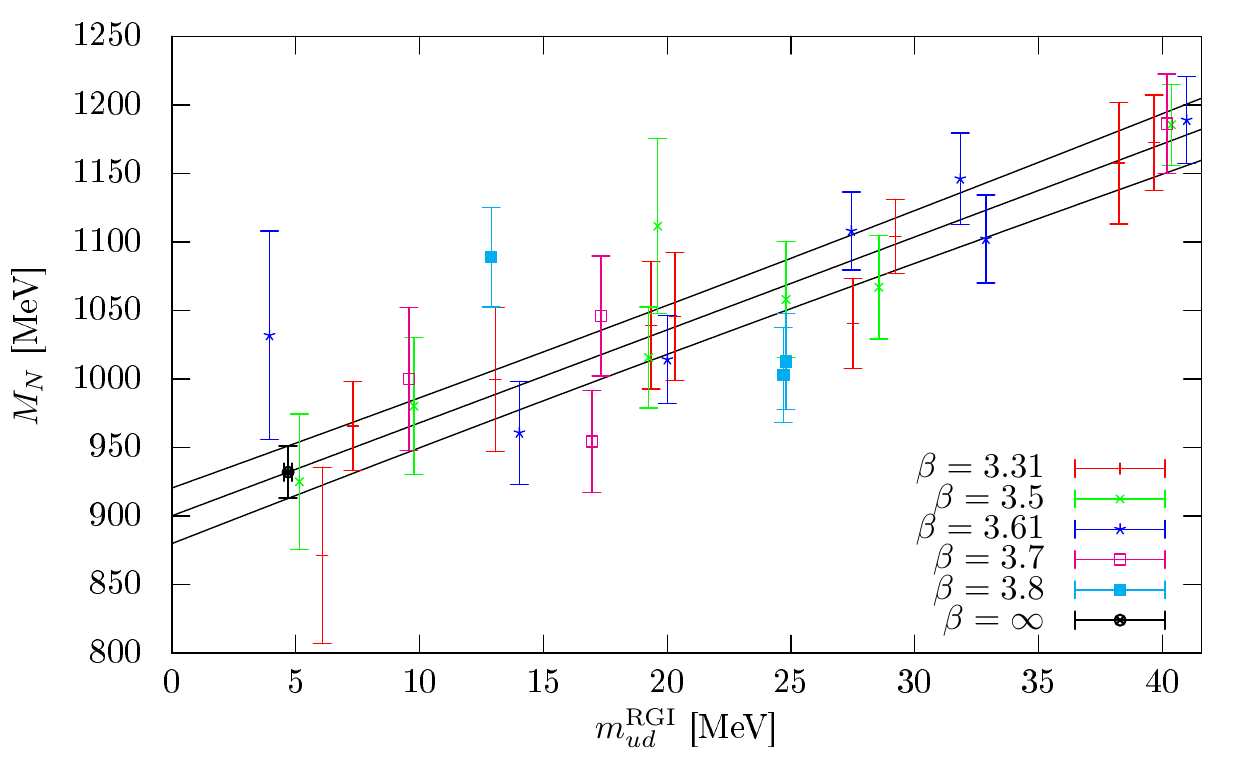}
\caption{$M_N$ plotted as a function of the renormalization group invariant (RGI) quark mass $m_{ud}^{(RGI)}$. Different colors and symbols denote each a different lattice spacing. The point in black denotes the value of $M_N$ at the physical point. The central curve corresponds to the best fit and the two other curves delimit the pointwise $68\%$ confidence interval. At a given value of $m_{ud}^{(RGI)}$, the fit has been used to shift the lattice results to the physical value of the other parameters (e.g., $m_s\rightarrow m_s^{(\Phi)}$, $a\rightarrow 0$, $\ldots$).}
\label{Fig. 2}
\vspace*{-0.45cm}
\end{figure}

\section{Preliminary results and outlook}

In order to assess the accuracy of the analyses, a good indicator is given by the fitted value for the mass of the nucleon $M_N$. The result that we obtain is $M_N = 957(22)(5)\ \mev$, where the first and second number in brackets correspond to the statistical and systematic error: the experimental value ($938.9\ \mev$) is safely recovered within errorbars.\\ 
\indent Figs. $\!2$ and $3$ show a typical dependence of $M_N$ on $m_{ud}$ and $m_s$, respectively, for one particular analysis. As expected, the slope in the strange-quark case has a large statistical error, as the error band in Fig. $3$ suggests.\\
\indent As for the preliminary results for the nucleon quark contents, they read $f_{udN} = 0.027(14)(4)$ and $f_{sN} =  0.18(8)(4)$. The $ud$ content is slightly lower than other $N_f\ge 2+1$ lattice results shown in Fig. $\!1$, but consistent within errors; on the other hand, $f_{sN}$ is systematically larger, but errorbars\nopagebreak\ are still large on this quantity.\\
\indent We wish to emphasize that our results are the only ones obtained from simulations with pion masses all the way down to its physical value, and even below. This means that the potentially large model-dependence associated with the extrapolation in $m_{ud}$ required in other calculations becomes a controlled interpolation error here. This is particularly important for the $ud$ content, which is related to the slope of $M_N$ with respect to
$m_{ud}$ at the physical value of $m_{ud}$. Indeed, comparing the present result for $f_{udN}$ to the one obtained in \cite{Durr10}, with simulations going down to $M_\pi\sim190\,\mev$, we find that the systematic error got quite reduced in the new calculation. A study of the various sources of systematic errors confirms that this difference comes from the uncertainty associated with reaching the physical $m_{ud}$ point.\\
\indent As seen in Fig. $\!3$, the dependence of $M_N$ on $m_s$ is small. Thus, the strange content of the nucleon remains a quantity which is difficult to determine, even with results all the way down to physical $m_{ud}$. A more detailed investigation shows that the main source of systematics is the possible contamination by excited states in the time correlator fits. Performing the analysis only on the $48$ procedures featuring a conservative choice for the fitted time ranges produces $f_{sN} = 0.199(96)(2)$ while the $48$ strategies with a more aggressive choice for this range result in $f_{sN} = 0.084(67)(5)$. Bootstrapping the statistical and systematic error in the difference yields $\Delta f_{sN} = 0.112(117)(1)$. 
Here improvement will only come with significantly more statistics and an increased lever arm in
$m_s$.


\begin{figure}
\includegraphics[width=0.85\textwidth]{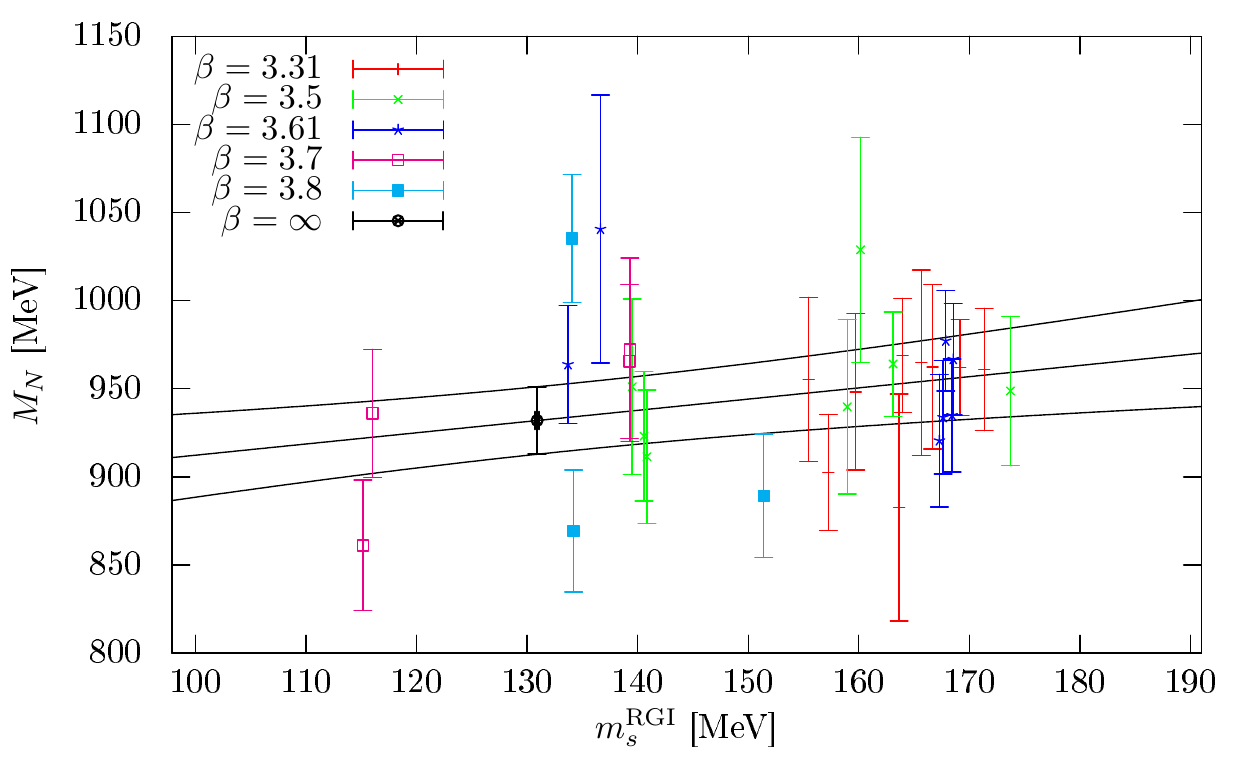}
\caption{Same as Fig. $\!2$ but for the $m_s^{RGI}$ dependence of $M_N$.}
\label{Fig. 3}
\end{figure}

\acknowledgments

Computations were performed using HPC resources provided by
GENCI-[IDRIS] (grant 52275) and FZ J\"ulich. This work was supported
in part by the OCEVU Labex (ANR-11-LABX-0060) and the A$^\star$MIDEX
project (ANR-11-IDEX-0001-02), funded by the "Investissements d'Avenir"
French government program and managed by the ANR, by CNRS grants GDR
$n^o$2921 and PICS $n^o$4707, by EU grants FP7/2007-2013/ERC 208740
and MRTN-CT-2006-035482 (FLAVIAnet), and by DFG grants FO 502/2, SFB-TR
55.


\begin{thebibliography}{99}


\bibitem{Gasser91} J.~Gasser, H.~Leutwyler and M.~Sainio,  
\emph{Sigma Term Update},
\emph{Phys. Lett.} {\bf B253} (1991) 252.

\bibitem{Pavan02} M.~Pavan, I.~Strakovsky, R.~Workman and R.~Arndt,  
\emph{The Pion nucleon Sigma term is definitely large: Results from a G.W.U. analysis of pi nucleon scattering data},
\emph{PiN Newslett.} {\bf 16} (2002) 110 [{\tt hep-ph/0111066}].


\bibitem{Alexandrou14} C.~Alexandrou et al. ,  
\emph{Baryon spectrum with $Nf=2+1+1$ twisted mass fermions},
\emph{Phys. Rev.} {\bf D90} (2014) 074501 [{\tt hep-lat/1406.4310}].

\bibitem{Ren14} X.-L.~Ren et al. ,  
\emph{An accurate determination of octet baryon sigma terms},
[{\tt hep-ph/1404.4799}].

\bibitem{Durr12} S.~D\"urr et al.,
\emph{Sigma term and strangeness content of octet baryons},
\emph{Phys. Rev.} {\bf D85} (2012) 014509 [{\tt hep-lat/1109.4265}].




\bibitem{Durr10} S.~D\"urr et al.,
\emph{Lattice QCD at the physical point: Simulation and analysis details},
\emph{JHEP} {\bf 1108} (2011) 148 [{\tt hep-lat/1011.2711}].

\bibitem{Borsanyi14} S.~Borsanyi et al.,
\emph{Ab initio calculation of the neutron-proton mass difference},
[{\tt hep-lat/1406.4088}].


\end{thebibliography}
\end{document}